\documentclass[twocolumn,showpacs,showkeys,preprintnumbers,amsmath,amssymb,amsfonts,euscript,revsymb,rmp]{revtex4}
\usepackage[dvips]{graphicx}
\usepackage{hyperref}
\usepackage{dcolumn}
\usepackage{bm}

\expandafter\ifx\csname package@font\endcsname\relax\else
 \expandafter\expandafter
 \expandafter\usepackage
 \expandafter\expandafter
 \expandafter{\csname package@font\endcsname}%
\fi

\begin{document}

\title{Many-Particle Quantum Cosmology}

\author{L.A. Glinka}
\email{laglinka@gmail.com, glinka@theor.jinr.ru}
\affiliation{Bogoliubov Laboratory of Theoretical Physics, Joint Institute for Nuclear Research, 141980 Dubna, Russia}
\date{\today}

\begin{abstract}
The Einstein--Friedmann Universe as whole quantum object can be treated as bosonic string mass groundstate, called a \textit{tachyon}, having negative mass square and a speed more than the speed of light. I present a brief review of results obtained from this point of view called Many-Particle Quantum Gravity approach - the monodromy problem in the Fock space, thermodynamics of the Universe, and the extremal tachyon mass model.
\end{abstract}


\maketitle
\section{Introduction}
The crucial source of knowledge about Nature is our Universe. Experimental data show both beautiful and mysterious structure of observed regions of the Cosmos. First of all we have to deal with very subtle the Large Scale Structure of Universe that up to today seems unexplained from theoretical physics point of view. The second problem is the Dark Matter existence and explanation of observational properties of light in the Universe in this context. An the last problem is the light that shows fascinating face in form of anisotropies of the Cosmic Microwave Background radiation. Both these facts and many others have unknown origin and theoretical elucidation.

The main goal of the present paper is an explanation of the leading results arose from the Many-Particle Quantum Gravity approach to the Einstein--Friedmann Spacetime, that can be called \emph{the Many-Particle Quantum Cosmology}. As it will be shown, the string theory context of this approach creates an elegant image of the Universe as the bosonic string mass groundstate. This fundamental state called \textit{the tachyon} is a physical object having negative mass square and a linear speed more than the speed of light $c$. In the case of Cosmology the tachyon has a mass that depends on the Friedmann scale factor, but locally the mass is spacetime constant. I begin form very concisely discussing of the first and the second quantization of the Universe modeled by the Einstein--Friedmann Spacetime, and as the main consequence of this way I point out the problem of monodromy in the Fock space of creation and annihilation operators. This approach creates an opportunity to constructive formulating of thermodynamics for the tachyon that is very briefly discussed in this paper. Finally, I consider a minimal model for the Many-Particle Quantum Cosmology - \emph{the extremal tachyon mass model}.

\section{Classical field theory}
Standard Cosmology can be formulated by Dirac--Arnowitt--Deser--Misner approach \cite{d,adm} to the Einstein--Hilbert action\footnote{The units $\hbar=c=k_B=\dfrac{8\pi G}{3}=1$ are used in this paper.} \cite{h,e1}
\begin{equation}\label{eh}
\mathit{S}_{\mathrm{EH}}=\int d^4x\sqrt{-g}\left(-\dfrac{1}{6}\mathcal{R}+\mathcal{L}\right),
\end{equation}
for a case of homogenous, flat, and isotropic model of the Universe given by the Einstein--Friedmann metric \cite{e,f}
\begin{equation}\label{ef}
ds^2=a^2(\eta)\left[(d\eta)^2-(dx^i)^2\right],~~d\eta=N(x^0)dx^0.
\end{equation}
Here $\mathcal{L}$ is the Lagrangian of all physical fields in the Universe, $\mathcal{R}$ is the scalar Ricci curvature, $g$ is a metric determinant, $\eta$ is the Dirac conformal time on the given Riemannian sufrace, $x^{\mu}$ $(\mu=0,1,2,3)$ are spacetime coordinates, $a$ is the Friedmann conformal scale factor, and $N$ is the lapse function in the Arnowitt--Dseser--Misner formalism. In result the metric (\ref{ef}) is equivalent to the constraints
\begin{equation}\label{c}
p_{a}^{2}-4V_0^2\rho(a)=0,
\end{equation}
where $p_{a}=-2V_0\dfrac{da}{d\eta}$ is canonical momentum conjugated to the Friedmann scale factor $a$, $V_0=\int d^3x$ is finite space volume, and 
\begin{equation}
\rho(a)=\dfrac{a^4}{V_0}\int{d^{3}x}~\mathcal{H}(x),
\end{equation}
has a sense of all physical fields energy distribution in the Universe multiplied by $a^4$, where $\mathcal{H}(x)$ is the Hamiltonian of these fields. The formal indentification
\begin{equation}\label{cor}
m^2\equiv-4V_0^2\rho(a),
\end{equation}
transforms the Dirac constraints (\ref{c}) into the form
\begin{equation}\label{c1}
p_a^2+m^2=0,
\end{equation}
that are the primary constraints for (free) bosonic string \cite{lt} with mass $m$. In considered case the string mass $m$ is dependent on the Friedmann scale factor $a$ that is degree of freedom of the Einstein--Hilbert theory (\ref{eh}) on the Riemannian surface given by the Einstein--Friedmann metric (\ref{ef}). However, it is global property of the mass, and locally, for given $a$ the tachyon mass has constant value. The square of mass (\ref{cor}) is negative and by this we have to deal with a tachyon, that is the bosonic string mass groundstate \cite{gsw,bn,p}. Using of Friedmann's definition of conformal time $\eta$ and Hubble's evolution parameter $H(a)$
\begin{equation}
d\eta=\dfrac{dt}{a(t)},~~H(a)=\dfrac{1}{a}\dfrac{da}{dt},
\end{equation}
where $t$ is the Friedmann cosmological time, leads to the relation
\begin{equation}\label{mass}
m=i|m|=i2V_0a^2H(a),
\end{equation}
with nontrivial condition given by the Hubble law
\begin{equation}
\mathcal{V}\int_{a^2}^{a_I^2}\dfrac{dy}{|m(y)|}=(t-t_I)^2,
\end{equation}
where $\mathcal{V}=V_0(t-t_I)$ is spacetime volume of given region of the Cosmos where the light comes to an observer, $y=a^2$ is integral variable, and index $I$ means an initial data. Direct applying of the canonical quantization by equal time commutation relations applying to the Dirac constraints (\ref{c}) leaves to the classical field equations in a form
\begin{equation}\label{ham1}
\dfrac{\partial}{\partial{a}}\left[\begin{array}{c}\Psi\\
\Pi_\Psi\end{array}\right]=\left[\begin{array}{cc}
0&1\\
m^{2}&0\end{array}\right]\left[\begin{array}{c}\Psi\\
\Pi_\Psi\end{array}\right],
\end{equation}
where $\Psi$ is the Wheeler--DeWitt wave function \cite{w1,w2}, and \mbox{$\Pi_\Psi$} is classical canonical momentum field conjugated to $\Psi$ understood as classical field. 
\section{Monodromy in the Fock space}
Conception of the wave function of the Universe as a solution of the Wheeler--DeWitt equation, obtained by first quantization and creating by compounding of equations (\ref{ham1}) was investigated by Hartle and Hawking in \cite{hh} and by Halliwell and Hawking \cite{hah}. This type considerations do not seem to give further physical results. I have proposed \cite{g, g1, g2} to consider \emph{the second quantization of two--component evolution} (\ref{ham1}). As it will be present here briefly, this procedure gives nontrivial results for physics of the Universe (\ref{ef}). In aim to the second quantization of (\ref{ham1}) we should introduce the field operator $\mathbf{\Psi}$ and $\mathbf{\Pi_\Psi}$ thats realize the bosonic canonical commutation relations
\begin{equation}
\left[\mathbf{\Pi_{\Psi}}[a],\mathbf{\Psi}[a']\right]=-i\delta_{aa'},\label{q}
\end{equation}
where $\delta_{aa'}\equiv\delta\left(a-a'\right)$, \mbox{$a\equiv a(\eta)$}, $a'\equiv a(\eta')$ for shortness, together with trivial commutators of two fields $\mathbf{\Psi}$, and two conjugate momenta fields $\mathbf{\Pi_\Psi}$. It is not difficult to check that the Von Neumann--Araki--Woods quantization \mbox{\cite{ccr1, ccr2}}
\begin{eqnarray}\label{2nd}
\left[\begin{array}{c}\mathbf{\Psi}[a]\\\mathbf{\Pi_\Psi}[a]\end{array}\right]
\!\!=\!\!\left[\begin{array}{cc}\dfrac{1}{\sqrt{2|m|}}&\dfrac{1}{\sqrt{2|m|}}\\
-i\sqrt{\dfrac{|m|}{2}}&i\sqrt{\dfrac{|m|}{2}}\end{array}\right]
\left[\begin{array}{c}\mathcal{G}[a]\\ \mathcal{G}^{\dagger}[a]\end{array}\right]\!\!,
\end{eqnarray}
realizes the general relations (\ref{q}) when and only when $\mathcal{G}[a]$ and $\mathcal{G}^{\dagger}[a]$ create bosonic type dynamical functional operator basis $$\mathcal{B}_{a}=\left\{\left[\begin{array}{c}\mathcal{G}[a]\\
\mathcal{G}^{\dagger}[a]\end{array}\right]\!:\!\left[\mathcal{G}[a],\mathcal{G}^{\dagger}[a']\right]\!=\!\delta_{aa'}, \left[\mathcal{G}[a],\mathcal{G}[a']\right]\!=\!0\right\}.$$ This basis can be treated as solution of quantized classical field theory (\ref{ham1}) equations
\begin{eqnarray}
\dfrac{\partial}{\partial{a}}\left[\begin{array}{c}\mathcal{G}[a]\\
\mathcal{G}^{\dagger}[a]\end{array}\right]=\left[\begin{array}{cc}
-m&\dfrac{1}{2m}\dfrac{\partial m}{\partial a}\\
\dfrac{1}{2m}\dfrac{\partial m}{\partial a}&m\end{array}\right]\left[\begin{array}{c}\mathcal{G}[a]\\
\mathcal{G}^{\dagger}[a]\end{array}\right].\label{fock}
\end{eqnarray}
Details \cite{g, g1, g2} show that for integrability of the quantum system (\ref{fock}) we must construct \emph{the Bogoliubov--Heisenberg basis}, obtained by the Bogoliubov automorphism in dynamical operator basis $\mathcal{B}_{a}$, and diagonalization of (\ref{fock}) to the Heisenberg evolution in assumed static operator basis $\mathcal{B}_0$
$$\mathcal{B}_{0}=\left\{\left[\begin{array}{c}\mathrm{w}\\
\mathrm{w}^{\dagger}\end{array}\right]: \left[\mathrm{w},\mathrm{w}^{\dagger}\right]=1, [\mathrm{w},\mathrm{w}]=0\right\}.$$
This basis has stable vacuum state and quantum field theory is well-defined. The basis $\mathcal{B}_0$ is related with $\mathcal{B}_{a}$ by the monodromy matrix $\mathbf{M}(a)$ in the Fock space
\begin{equation}\label{mon}
\mathcal{B}_{a}=\mathbf{M}(a)\mathcal{B}_0,
\end{equation}
with $\mathrm{\mathbf{M}}(a)$ as
\begin{equation}\nonumber
\left[\begin{array}{cc}
\Bigg(\sqrt{\bigg|\dfrac{m}{m_I}\bigg|}+\sqrt{\bigg|\dfrac{m_I}{m}\bigg|}\Bigg)\dfrac{e^{\lambda}}{2}\vspace*{5pt}&
\Bigg(\sqrt{\bigg|\dfrac{m}{m_I}\bigg|}-\sqrt{\bigg|\dfrac{m_I}{m}\bigg|}\Bigg)\dfrac{e^{-\lambda}}{2}\\
\Bigg(\sqrt{\bigg|\dfrac{m}{m_I}\bigg|}-\sqrt{\bigg|\dfrac{m_I}{m}\bigg|}\Bigg)\dfrac{e^{\lambda}}{2}&
\Bigg(\sqrt{\bigg|\dfrac{m}{m_I}\bigg|}+\sqrt{\bigg|\dfrac{m_I}{m}\bigg|}\Bigg)\dfrac{e^{-\lambda}}{2}\end{array}\right],
\end{equation}
where
\begin{eqnarray}\label{im}
\lambda=\lambda(a)=\pm\int_{a_I}^{a}m~da,
\end{eqnarray}
is integrated mass of considered free bosonic string. This well-defined quantum field theory description permits build formal thermodynamics for the tachyon. 
\section{The Tachyon Thermodynamics}
Presented way gives an opportunity to construct thermodynamics of the Universe modeled by the tachyon. In the static basis $\mathcal{B}_0$ we use description in the Gibbs ensemble, and one can compute formally all thermodynamics characteristics\footnote{By existing of stable vacuum state in the static basis we have to deal with thermodynamical equilibrium state of quantum states of considered system. Presented relations are result of the density functional method used for one-particle density functional.}
\paragraph{Occupation number $\mathrm{n}=\mathrm{n}(a)$}
\begin{equation}\label{n}
\mathrm{n}=\dfrac{1}{4}\left|\sqrt{\left|\dfrac{m}{m_I}\right|}-\sqrt{\bigg|\dfrac{m_I}{m}\bigg|}\right|^2,~~\langle\mathrm{n}\rangle=2\mathrm{n}+1
\end{equation}
\paragraph{Entropy $\mathrm{S}=\mathrm{S}(a)$}
\begin{equation}\label{ent}
\mathrm{S}=-\ln(2\mathrm{n}+1).
\end{equation}
\paragraph{Internal energy $\mathrm{U}=\mathrm{U}(a)$}
\begin{equation}\label{U}
\mathrm{U}=\left(\dfrac{1}{2}+\dfrac{4\mathrm{n}+3}{2\mathrm{n}+1}\mathrm{n}\right)|m|.
\end{equation}
\paragraph{Chemical potential $\mu=\mu(a)$}
\begin{equation}\label{mu}
\mu=\Bigg(1+\dfrac{1}{(2\mathrm{n}+1)^2}-\dfrac{1}{2}\dfrac{4\mathrm{n}+1}{4\mathrm{n}^2+2\mathrm{n}}\sqrt{\dfrac{\mathrm{n}}{\mathrm{n}+1}}\Bigg)|m|.
\end{equation}
\paragraph{Temperature $\mathrm{T}=\mathrm{T}(a)$}
\begin{eqnarray}\label{temp}
\mathrm{T}=\dfrac{1+\left(\dfrac{2\mathrm{n}}{2\mathrm{n}+1}\right)^2+\dfrac{8\mathrm{n}^2+8\mathrm{n}+1}{4\mathrm{n}+2}\sqrt{\dfrac{\mathrm{n}}{\mathrm{n}+1}}}{2\ln(2\mathrm{n}+2)}|m|.
\end{eqnarray}
These formal thermodynamics describes a thermal statistical mechanics properties of system of quantum states of the Universe modeled by the one quantum object, that is the tachyon being mass groundstate of bosonic string, in state of thermodynamical equilibrium guaranteed by proper choice of operator basis.
\section{The extremal tachyon mass model}
Now we are going to discuss some very special case of presented formalism - \emph{the extremal tachyon mass model}. This model springs from treating of the integrated mass of the tachyon (\ref{im})
\begin{equation}
\lambda=\lambda(a)=\int_{a_I}^a m(a)da,
\end{equation}
as the global Weyl characteristic scale of the Universe. This means simply that $\lambda$ fulfill the d'Alembert equation
\begin{equation}
\dfrac{\partial^2\lambda(a)}{\partial a^2}=0,
\end{equation}
which means a constant value of the tachyon mass given by an initial data
\begin{equation}\label{mm}
m(a)=m_I\equiv m(a_I).
\end{equation}
From the relation (\ref{mass}) arises a conclusion that the Hubble parameter in the extremal mass tachyon model is
\begin{equation}\label{hue}
H(a)=\dfrac{a_I^2}{a^2}H_I,
\end{equation}
where $H_I\equiv H(a_I)$ is an initial data of the Hubble evolution parameter. We will call (\ref{hue}) as \emph{the extremal Hubble parameter}.

Solution of the bosonic string constraints (\ref{c1}) for the extremal Hubble parameter (\ref{hue}) is
\begin{equation}
a(t)=a_I\sqrt{1+2H_I|t-t_I|},
\end{equation}
and in result the dependence between the scale $\lambda$ and cosmological time $t$ is as follows
\begin{equation}
\lambda(t)=m_Ia_I\left|\sqrt{1+2H_I|t-t_I|}-1\right|.
\end{equation}
For the extremal tachyon mass model, the operator evolution (\ref{mon}) has a very simple form
\begin{equation}\nonumber
\left[\begin{array}{c}\mathcal{G}(t)\\
\mathcal{G}^{\dagger}(t)\end{array}\right]=\left[\begin{array}{cc}e^{\lambda(t)}&0\\0&e^{-\lambda(t)}\end{array}\right]\left[\begin{array}{cc}\mathrm{w}\\
\mathrm{w}^{\dagger}\end{array}\right].
\end{equation}
As a consequence of the relation (\ref{mm}), which means that the averaged occupation number $\langle\mathrm{n}\rangle$ of quantum states of the tachyon in the second formula (\ref{n}) equals unity, arises that the temperature of the tachyon (\ref{temp}) is
\begin{equation}\label{temp1}
\mathrm{T}=\dfrac{|m_I|}{2\ln2}=V_0\dfrac{a_I^2H_I}{\ln2},
\end{equation}
in spite of a value of the Von Neumann-Boltzmann entropy (\ref{ent}) is zero. The extremal tachyon mass model is the minimal model for the quantum cosmology (\ref{mon}). General models can be consider as corrections to the extremal mass model. From the form of the extremal Hubble parameter (\ref{hue}) we conclude that the minimal model means presence of the Radiation only, and general models should studying the Matter and the Dark Matter contributions as corrections.
\section{The meaning of its all}
The first fundamental question is \emph{what is the Many-Particle Quantum Cosmology}? In this paper was presented a string theory face of this approach. Truly, presented approach formulates the Quantum Cosmology by dynamics of the tachyon being bosonic string mass groundstate. It is \emph{the correct} meaning of Many-Particle Quantum Cosmology.

The second problem is \emph{what is a crucial problem for correct formulation of Quantum Gravity}? Serving the analogy method to presented above approach, one can say that the problem lies in investigating of bosonic string dynamics and formulation of the second quantization of this classical field theory. By this the recept for the proper construction of Quantum Gravity is building of some string quantum field theory in the Fock space of creation and annihilation operators. It is the main problem for future considerations.

The author nurture hopes that Many-Particle Quantum Gravity approach will find further applications.
\section*{Acknowledgements}

The author got remarkable valuable inspirations by discussions with \mbox{B.M. Barbashov}, \mbox{I. Bia\l ynicki-Birula}, \mbox{S.J. Brodsky}, \mbox{S. Ferrara}, \mbox{G. $\acute{}$ t Hooft}, \mbox{L.N. Lipatov}, \mbox{V.B. Priezzhev}, \mbox{S. Pokorski},  and \mbox{D.V. Shirkov}.

\end{document}